# Voice EHR: Introducing Multimodal Audio Data for Health

James Anibal[1,2], Hannah Huth[1], Ming Li[1], Lindsey Hazen[1], Veronica Daoud[3*], Dominique Ebedes[3*], Yen Minh Lam[4], Hang Nguyen[4], Phuc Vo Hong[4], Michael Kleinman[5*], Shelley Ost[5*], Christopher Jackson[5*], Laura Sprabery[5*], Cheran Elangovan[5*], Balaji Krishnaiah[5*], Lee Akst[6,7*], Ioan Lina[7*], Iqbal Elyazar[8*], Lenny Ekwati[8*], Stefan Jansen[9*], Richard Nduwayezu[10*], Charisse Garcia[1], Jeffrey Plum[1], Jacqueline Brenner[1], Miranda Song[1], Emily Ricotta[11,12], David Clifton[2], C. Louise Thwaites[3], Yael Bensoussan[3], Bradford Wood[1]

[1]Center for Interventional Oncology, NIH Clinical Center, National Institutes of Health, USA
[2] Computational Health Informatics Lab, Oxford Institute of Biomedical Engineering, University of Oxford, UK
[3]Morsani College of Medicine, University of South Florida, Tampa, Florida, USA
[4]Oxford University Clinical Research Unit, Ho Chi Minh City, Vietnam
[5]College of Medicine, University of Tennessee Health Sciences Center, Memphis, Tennessee, USA
[6]Johns Hopkins Voice Center, Johns Hopkins University, Baltimore, MD, USA
[7]Department of Otolaryngology-Head and Neck Surgery, Johns Hopkins University School of Medicine, Baltimore, MD, USA
[8]Oxford University Clinical Research Unit Indonesia, Jakarta, Indonesia
[9]College of Medicine and Health Sciences, University of Rwanda, Kigali, Rwanda
[10]King Faisal Hospital, Kigali, Rwanda
[11]Epidemiology and Data Management Unit, National Institute of Allergy and Infectious Diseases, Bethesda, MD, USA
[12]Department of Preventive Medicine and Biostatistics, Uniformed Services University
*Contributed equally

**Abstract**

Artificial intelligence (AI) models trained on audio data may have the potential to rapidly perform clinical tasks, enhancing medical decision-making and potentially improving outcomes through early detection. Existing technologies depend on limited datasets collected with expensive recording equipment in high-income countries, which challenges deployment in resource-constrained, high-volume settings where audio data may have a profound impact on health equity. This report introduces a novel data type and a corresponding collection system that captures health data through guided questions using only a mobile/web application. The app facilitates the collection of an audio electronic health record ("Voice EHR") which may contain complex biomarkers of health from conventional voice/respiratory features, speech patterns, and spoken language with semantic meaning and longitudinal context – potentially compensating for the typical limitations of unimodal clinical datasets. This report presents the application used for data collection, initial experiments on data quality, and case studies which demonstrate the potential of voice EHR to advance the scalability/diversity of audio AI.

## 1. Introduction

The COVID-19 pandemic underscored the limitations of healthcare systems and highlighted the need for data innovations to support both care providers and patients. The high volume of patients seeking medical care for COVID-19 and other viral infections has caused extraordinary challenges, including long waitlists, limited time for each patient, increased testing costs, exposure risks for healthcare workers, and documentation burdens.[1] Adding to the problem, the world is facing nursing and physician shortages which are expected to rise dramatically over the next 10 years.[2-4] This contributes to the increasing rates of provider burnout and a loss of trust in the healthcare system, both of which have been particularly severe since the onset of the COVID-19 pandemic.[5-7] To address these problems, artificial intelligence (AI) has been proposed as a mechanism to rapidly perform key clinical tasks such as diagnostics, triage, and patient monitoring, improving the efficiency of the healthcare system. This has become particularly true with the advent of GPT and other multimodal large language models (LLMs), which have advanced capabilities in question answering, image interpretation, programming, and other complex tasks.[8-9] As a result, technology companies have begun to develop foundation AI models for the healthcare space. However, these are mainly designed for processing and diagnostic tasks with privileged data (e.g., images) or as a Chatbot tool for general question-answering rather than providing specific recommendations from patient data.[10-11] While future LLMs may add value to the healthcare system, serious data challenges remain for the widespread, equitable deployment of AI models in healthcare. Below, several primary obstacles are outlined:

**Data Availability and Interoperability:** In many cases, clinical AI models require correlated data – information from different sources related to the same patient within the same approximate period of time. Datasets also require extensive curation, which is often expensive, inconvenient, and frequently overlooked as a challenge in the development of health AI. Multimodal data must be linked from across disjointed sources/centers, which often have incompatible systems and different regulatory structures.

**Excluding underserved groups:** Currently, many AI technologies are dependent on the availability, quality, and breadth of data in electronic health records (EHR), which are often unavailable or inaccurate in many settings, particularly in resource-constrained areas such as low- and middle- income countries (LMICs) or rural areas in high-income countries.[12] These disparities are due to many factors, which include biased allocation of healthcare services, gaps in insurance coverage, and other barriers (e.g., transportation) due to a lack of providers or facilities.[13] As a result, training data for AI models is often biased against underserved populations.[14]

**Misalignment with clinical processes:** The data collected in current clinical workflows is incompatible with most AI systems, causing development challenges and hesitancy from healthcare workers, who make decisions based on patient reporting, their own observations, and various tests - not narrow unimodal datasets collected in research settings. Figure 1 (below) highlights the disconnect between the conditions for data collection in funded research projects and those for data collection or inference in real-world settings, which, in some ways, are more similar to the uncontrolled nature of data mined from online sources.

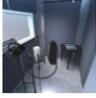

**Figure 1:** Comparison of audio data collected from research studies with hospital audio data and data from public online sources (i.e., YouTube).

**Contributions**

This work makes the following contributions to the structure and collection of healthcare data:

1. Development of an online application (Healthcare via Electronic and Acoustic Records, "HEAR") to facilitate the collection of semi-structured multimodal data (text-audio pairs) for developing AI models. The application is designed to be intuitive for patients and technically lightweight for deployment in low-connectivity areas. This system simultaneously captures patient-reported health information (via recorded speech) and unique variations in sound data (changes in voice/speech) without the potential inconsistency of methods such as ambient listening. In a single setting, the HEAR application facilitates rapid collection of health data for training AI models, including retrospective context and factors related to circumstances/lifestyle. The user is not required to type text into a lengthy form, which may cause respondent fatigue and result in data with decreasing accuracy and increasing missingness.[15,16]

2. Presentation of demographic statistics, experimental results, and case studies from an initial Voice EHR dataset. Experiments were run using GPT-4o to compare the robustness of information contained in the Voice EHR with data collected via manual inputs. These preliminary results demonstrate the potential viability of low-cost voice EHR data collected across multiple settings, including hospitals.

## 2. Related Work

Audio data has previously shown potential as a diagnostic tool. The idea that patients with certain conditions might present with unique changes in their voice before showing more progressive signs of disease largely originated with Parkinson's disease. Multiple studies have shown that Parkinson's disease is associated with characteristic and progressive changes in phonation over the disease course, including biomarkers such as decreased word stress, softened consonants, abnormal silences, and monotone speech.[17-20] Similarly, many studies have since identified specific voice changes in patients with asthma, COPD, interstitial lung disease, rheumatoid arthritis, chronic pain, diabetes, and laryngeal cancer.[21-27] The formation of multi-site data generation projects like the Bridge2AI Voice Consortium shows the increasing interest in leveraging voice as a data modality for healthcare applications.[28]

During the COVID-19 pandemic, the demand for low-cost digital healthcare solutions surged, providing an ideal setting to advance audio AI technology. Multiple methods were developed which allowed for the prediction of COVID-19 status (positive/negative) or variant status via machine learning models trained to detect differences in acoustic features from voice.[29-42] However, many of these models were not deployed due to limitations of the training datasets (described below), and there was no significant evidence that voice/audio AI methods improved COVID-19 screening during the pandemic.[43-44]

1. **Dataset Size/Diversity:** Many voice AI studies are reliant on small datasets collected from a narrow range of English-speaking patients using high-cost technology like recording booths, preventing deployment in hospitals or at-home settings (See Figure 1 for comparison of "research data" and data from real-world environments).

2. **Data Quality**: Many studies were built around crowdsourced datasets, which face significant issues with data quality – reliable annotations (specific indications of disease or health state) are difficult to achieve when collecting limited data from a wide range of possible environments.[45-46] Many datasets, which contain scripted voice samples, will not be usable due to the lack of context needed to account for sources of noise. Moreover, very few datasets were curated through partnerships with healthcare workers in clinical settings, and, as such, do not confirm diagnosis of COVID-19 or other illnesses.

3. **Data Breadth:** Past audio AI studies, particularly those involving COVID-19 screening/diagnostics, often excluded patients with confirmed cases of other respiratory illnesses – in some cases, only healthy samples were included in the control cohorts.[39-42] A typical user of any testing method would do so because of newly emerging symptoms, which may then confuse an AI model which was trained only to separate between a disease state and fully healthy controls or chronic conditions. More specifically, although illnesses like COVID-19 can cause laryngitis and inflammation of the vocal cords causing voice changes, many other factors, such as smoking habits, can also cause laryngitis.[47]

This study introduces "Voice EHR" – patients share their past medical history and progression of present illness through audio recordings, creating a patient-driven temporal record of clinical information to contextualize breathing, voice, and speech data collected simultaneously.

## 3. Methods

The development of AI models to accurately detect audio biomarkers of disease is dependent on the acquisition of robust training datasets from diverse settings. The proposed "Voice EHR" methods were designed to enable semantic representations of clinical information containing approximate temporal context (e.g., changes from baseline health) with correlated samples of acoustic data: voice/breathing sounds and speech patterns.

This study was approved by the Institutional Review Board of the U.S. National Institutes of Health (NIH). Informed consent was obtained from all participants prior to data collection, using a built-in consent form on the online application. Data is stored on NIH-secured cloud servers maintained by Amazon Web Services (AWS).[48] No personally identifying information is stored at this time.

### 3.1 Participant Recruitment and Study Population

The data collection process was deployed through two primary channels: 1) public use of the application, which is available online at [www.hearai.org](www.hearai.org), and 2) partnerships with healthcare professionals working at collaborating point-of care settings. The HEAR app is low-cost, low-bandwidth, fast/easy to use, and does not rely on any specific expensive technologies (e.g., recording booths), facilitating partnerships with healthcare workers. Collaboration with healthcare professionals will help improve the reliability of voice EHR data by providing validated annotations through recruitment of patients with confirmed diagnoses. Providers may engage with patients and ask follow-up questions during the collection process if necessary to enhance data robustness or if internally useful within the clinical workflow (this can be removed before analysis of the sound data). The application can be used by both providers and patients.

### 3.2 Data Collection

The HEAR application was designed to efficiently collect multimodal audio data for health – voice EHR – via a combination of short survey questions and recorded voice/speech/breathing tasks. The HEAR app contains three main sections (Fig. 2 – left). After obtaining informed consent, data collection begins with multiple-choice questions focused on basic health information (pages 1 – 5). This section is necessary during the data collection process to ensure a balanced training dataset for initial model development and validation. The recorded Voice EHR data is collected based on written instructions (pages 7 – 12). The final section (also audio data) is completed with the assistance of a researcher or care provider to document findings, next steps, diagnosis, and other components of the appointment (pages 14 – 17). Control participants do not complete pages 4, 8, or 14-17. For this study, a control is defined as a participant who does not have an illness which is known to affect the voice or the respiratory system.

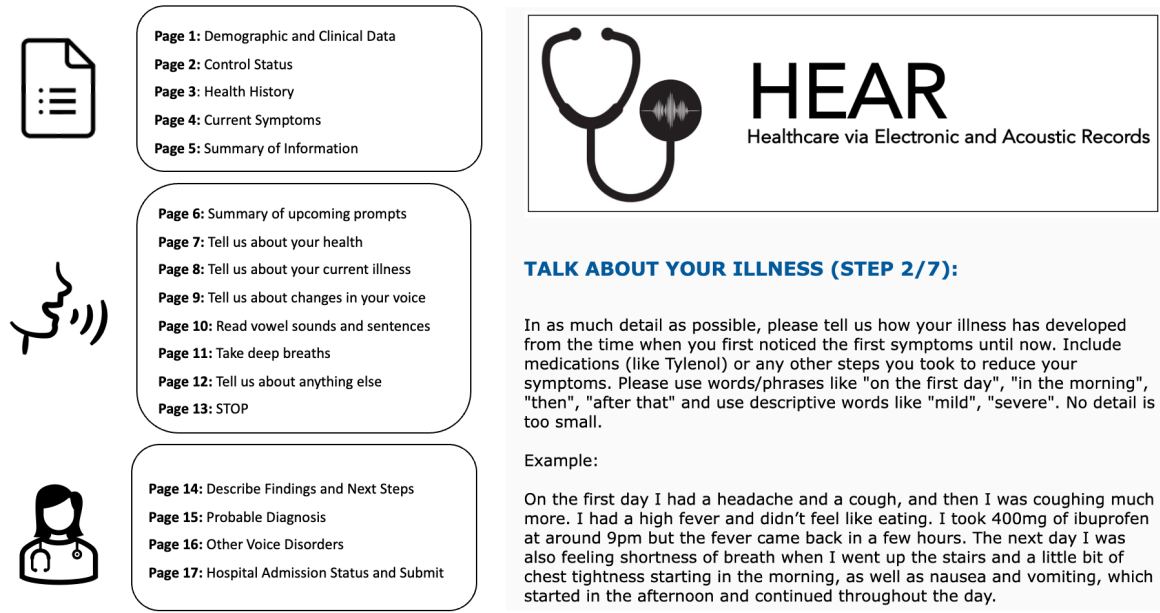

**Figure 2. (Left)** Overview of the voice EHR data collection app, including initial survey, patient audio, and information from HCWs. **(Right)** Screenshot from the app (second audio prompt).

## 3.3 Audio Data

This section of the report describes methods for collecting multimodal audio data, containing information on voice/breathing sounds and speech patterns as well as semantic meaning from spoken language. Each prompt is designed based on real-world clinical workflows, which may enable the collection of training data that is more aligned with existing healthcare systems. Table 1 contains the voice prompts and a short descriptor of each. After collection, all audio recordings containing semantic meaning were transcribed. Automated speech recognition for this study was done with Whisper, a top-performing foundation model for speech-to-text tasks.[49]

**Table 1:** Participant Prompts included on the HEAR Application for audio data collection.

| Prompt | Purpose | Completed By |
|---|---|---|
| Please tell us background information about your health before your current illness, including: Chronic conditions (such as high blood pressure or diabetes), Recent illnesses (for example, COVID-19), Other physical health problems, Mental health problems, such as anxiety, Medications you currently take, Any recent changes to your medication which made you feel differently. | Establishes a baseline to contextualize changes due to illness, either in sounds, speech patterns, or spoken words. | Patients, Controls |
| In as much detail as possible, please tell us how your illness has developed from the time when you first noticed symptoms until now. Include any medications you took (like Tylenol) or steps you use to reduce your symptoms. Please use words/phrases like "on the first day", "in the morning", "then", "after that" and use descriptive words like "mild", "severe". No detail is too small. (3 min.) | Captures the complaint of the patient by approximating a record of illness progression. | Patients Only |

| Prompt | Purpose | Participants |
|---|---|---|
| Please tell us if you or anyone else has noticed any recent changes in your voice (like hoarse, raspy, or lost voice) speech (like difficulty getting words out or slurring words), or breathing. If so, describe these changes. These should be changes that started around the same time as this illness episode, not any chronic long-term changes. (1 min.) | Establishes an "audio" baseline to contextualize changes in voice/speech which may arise from lifestyle factors/past conditions or may be a biomarker of disease. | Patients, Controls |
| **Part 1:** Say each of these vowels for as long as you can. aaaaa (as in *made);* eeeee *(beet);* ooooo *(cool)* <br><br> **Part 2:** Read these sentences: "When the sunlight strikes raindrops in the air, they act as a prism and form a rainbow. The rainbow is a division of white light into many beautiful colors. These take the shape of a long round arch, with its path high above, and its two ends apparently beyond the horizon." | Conventional voice and respiratory data for analysis of sound changes. | Patients, Controls |
| **Part 1:** Hold the device near your nose and record yourself breathing normally for 30 seconds with your mouth closed. <br><br> **Part 2:** Hold the device near your mouth and record yourself taking 3 deep breaths through your mouth. | Conventional respiratory data for analysis of breathing changes and determination of respiratory rate. | Patients, Controls |
| Is there anything else you would like us to know about your health or circumstances that you feel we have missed? <br> For example, you can tell us about: your employment, your lifestyle habits, and/or any challenges you have had with the healthcare system, including delays with receiving care or problems with quality of care that may have impacted your health. | Captures specific circumstances related to health which the patient considers to be important. | Patients, Controls |
| Your physician or other provider should briefly describe the physical exam (given to you by the physician), any available lab results, imaging studies, the diagnosis, and other next steps related to testing, treatment, or monitoring the illness. If the healthcare provider is not available or you are at home, you can record this information yourself. | Audio approximation other types of multimodal data which may be key context for the patient data collected by the application. | Patients Only |

### *Initial Inputs: Demographic and Clinical Information for Data Annotation*

AI models developed from voice EHR data may be trained to perform clinical tasks using only multimodal audio data. However, in the experimental stages, respondents are asked to complete an initial set of questions to contextualize the collected audio data. This is done to ensure class balance, account for possible sources of bias, and run comparative experiments. These data include race, sex, symptoms (including duration and progression), education, insurance, and health history. Zip codes are also collected for epidemiological studies.

### *Semi-structured Audio Data: Voice EHR Prompts*

*Prompt 1: Health baseline*

The health baseline prompt is designed to provide background data on the participant, ensuring that disease can be modelled as a function of change from a fixed point. Purely cross-sectional datasets are unrealistic, potentially misinforming clinicians in real-world scenarios: no patient would be seen, let alone treated, before the care team reviewed the medical records or collected past medical history.

*Prompt 2: Illness Trajectory*
The second prompt is designed to capture a key interaction between a patient and their provider: "What brings you in today?". During this interaction, temporal descriptions of illnesses and corresponding patient-initiated interventions (e.g., "taking Tylenol") are collected, mirroring basic clinical assessments. The aim of this prompt is to ensure clinical information with temporal context is available to complement the sound data. The application asks patients to use basic terminology to describe, in chronological order, the progression of their illness with any associated signs, symptoms, complications, and corresponding interventions. Collecting this information through an audio recording is less burdensome than a typed/written form, may contain more detailed information, could potentially serve as a substitution for conventional time-series EHR data which is often sparse or unavailable, especially regarding over-the-counter and alternative therapies.

*Prompt 3: Voice baseline*
Past Voice AI studies have shown the obstructive impact of variables such as chronic laryngeal conditions or lifestyle factors such as smoking.[35] As such, the HEAR application prompts the patient to report any recent changes in voice, speech, or breathing noticed by themselves or others. As with Prompt 1, this prompt aims to replace baseline information in original form (e.g., voice samples from prior to illness), which may be unobtainable for ill patients using the system for the first time. This data may reduce the confounding effect of voice sounds or speech patterns which do not correlate to the current complaint.

*Prompt 4: Conventional Acoustic Data*
Prompt 4 facilitates the collection of conventional acoustic data that is often used in voice AI studies. The first task (Prompt 4, part 1), in which the patient phonates an elongated vowel for as long as possible, may help assess the impact of different variables on how air flows over the vocal cords and indicate the current overall function of the respiratory system. This prompt is a simple method of collecting acoustic features, can be easily translated into other languages, and has been previously used in projects involving crowdsourced data.[35] Prompt 4, part 2 – the "rainbow passage" – is a validated passage designed to maximize the diversity of acoustic features contained in a single data sample, ensuring that biomarkers are not missed due to limited/narrow inputs.[50] These data are collected not only to ensure that pure sound samples are available alongside the free speech with semantic meaning, but also as a mechanism for interoperability and comparison with unimodal data from past studies.

*Prompt 5: Conventional Breathing Data*
Participants are asked to breathe through the nose normally for 30 seconds (prompt 4, part 3) and take 3 deep breaths with the mouth open (prompt 4, part 4), facilitating tasks such as the calculation of respiratory rate (widely used in continuous vital sign monitoring) and the capture of dangerous airway conditions such as stridor or distinctive alterations from supraglottic edema.[51-52]

*Prompt 6: Additional Information*
To further ensure that Voice EHR data contains patient-centered data about medical history and the present illness, prompt 6 asks if the respondent has any other information that may be important to share (i.e., any contributing information that might not have been covered by past prompts), including challenges faced in engaging with the healthcare system. The addition of this information may lead to improvements in model performance when compared to past health datasets, which have been biased against underserved minority groups or individuals with unique clinical needs not considered in the design of structured EHR systems and standardized surveys.

*Prompt 7: Diagnosis and Treatment Plan*
If available, a healthcare worker will be asked to provide a brief recorded description of the appointment, diagnosis, and treatment plan. This recording may approximate types of clinical data which are often not collected/stored in low-resource settings, such as diagnostic tests or other lab results.

### 4. Preliminary Results

### 4.1 Dataset Statistics
This study resulted in the development of an application for the collection of multimodal audio data. Initial data collection work was focused on (1) illnesses which are known to impact the respiratory system and (2) the collection of diverse controls, defined as any data which does not fall into the first category. These early efforts resulted in a multi-site dataset of 130 English-speaking patients.

The total combined length of the recordings was 5.3 hours. If a participant recorded fewer than 2 audio samples, provided low-quality/noisy audio samples which could not be transcribed, or did not complete the demographic/clinical information sections (Pages 1-5), that data was excluded from this study.

Figure 3 presents demographic statistics for the patients in the initial dataset, including race, age, gender identity, and location of recording (hospital/clinic, home, other). Notably, in contrast to other crowdsourced/multi-site voice data generation projects, over half of the samples came from hospital settings. [28,45]

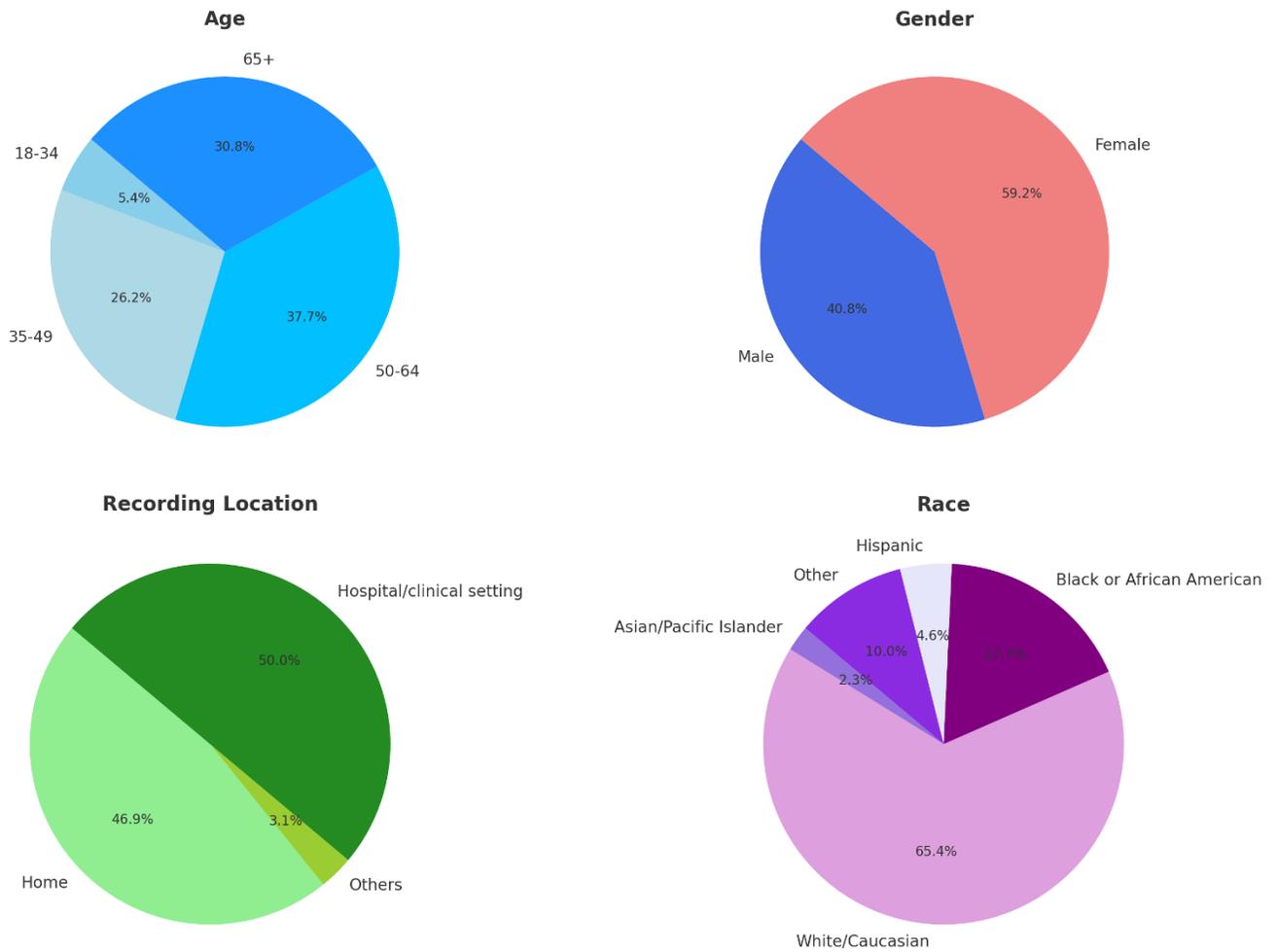

**Figure 3:** Demographic statistics for the initial voice EHR dataset (n=130)

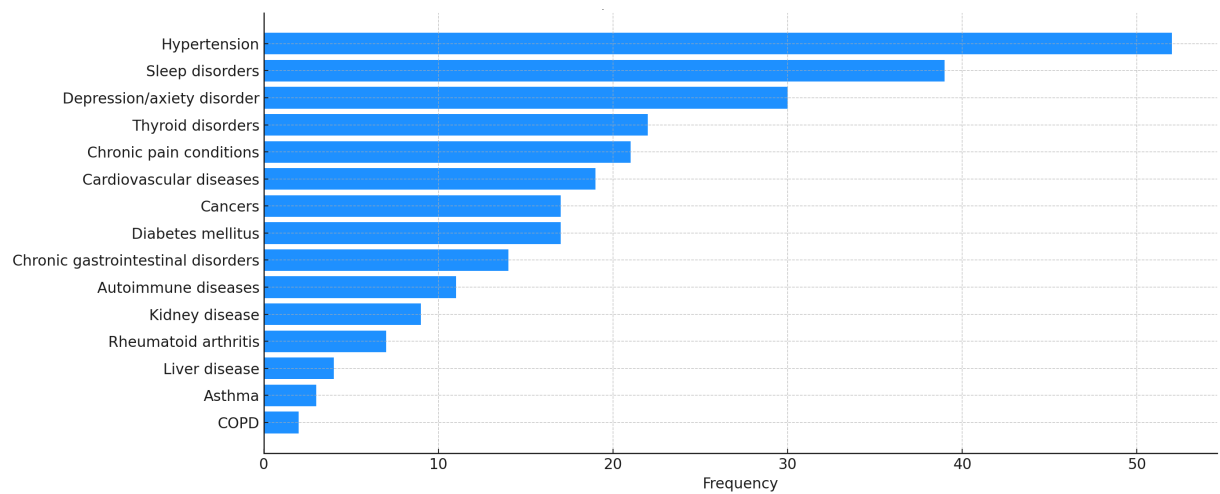

**Figure 4:** Occurrences of different health conditions within the initial Voice EHR dataset.

Figure 4 shows the prevalence of health conditions in the dataset, indicating a high occurrence of chronic conditions like hypertension, sleep disorders, depression/anxiety, thyroid disorders, and pain conditions.

## 4.2 Viability of Voice EHR Data

Large language models (LLMs) were used for conducting additional experiments to compare the information contained in Voice EHR audio recordings with the data initially provided by the patient through manual methods (e.g., multiple choice, short answer). These experiments were performed using Voice EHR from participants with an illness who completed the prompts related to health history and current complaint (Prompts 1-2), resulting in a subset of 41 data points.

In the experiments, GPT-4o was instructed to compare the audio transcripts with the manually input information and rate the two data sources based on a simple rubric designed to reflect general utility of the data in a potential healthcare assessment (Figure 5, Left).[53] The manually input information included the following variables: co-morbidities/health challenges (select all that apply, short answer), current symptoms (select all that apply, short answer), progression of symptoms (multiple choice), and duration of symptoms (multiple choice). Results of this experiment showed that, despite averaging less than 90 seconds in length, audio data on health history and current complaint was consistently more informative in the context of a health assessment (Figure 5, Right).

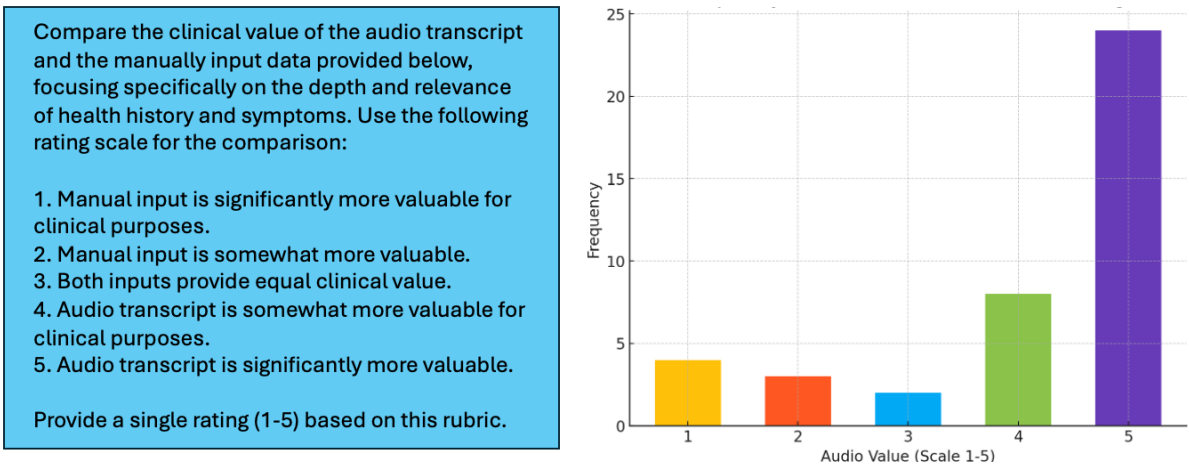

**Figure 5:** <u>Left:</u> Prompt used for instructing GPT-4o to compare Voice EHR audio transcripts with variables collected through conventional mechanisms on the HEAR application. <u>Right:</u> Distribution of LLM-generated ratings for the 41 patients included in this experiment.

Figure 5 (Right) shows that, in 83% of cases, the semi-structured audio data from the HEAR application was equally or more informative than manually input data (rating > 2). In 59% of cases, the Voice EHR audio transcript was significantly more valuable (rating=5). The mean LLM-generated rating was 4.10, with a median of 5 and a standard deviation of 1.36.

## 4.3 Case Studies of Initial Data

To further demonstrate the potential value of information contained in Voice EHR data, examples of basic health information and audio transcripts for patients with illnesses and control participants are presented in Tables 2-6. This is a limited sample of the dataset, for illustrative purposes.

**Table 2:** Examples of Basic Health Information from the HEAR application.

|  | **Patient A** | **Patient B** | **Patient C** | **Control A** | **Control B** | **Control C** |
|---|---|---|---|---|---|---|
| **Age** | 40 | 55 | 74 | 52 | 75 | 56 |
| **Weight** | 175 | 117 | 152 | 155 | 175 | 139 |
| **Sex** | Male | Female | Female | Female | Female | Female |
| **Race** | White | White | Hispanic | No Response | White | Black/AA |
| **Occupation** | Physician | Nurse | Nurse | Nurse | Retired | Landscaper |
| **Insurance** | Private | Public | Private | Public | Public | Private |
| **Education** | Graduate | College | Graduate | College | Graduate | College |
| **Recording** | Home | Home | Hospital | Home | Home | Home |
| **Health History** | None | None | Hypertension Cardiovascular disease Thyroid disease | Chronic pain Autoimmune Sleep disorders Depression | Thyroid Disorders Cancer Sleep disorders | MS Cancer |
| **Symptoms** | Cough Sore throat | Headache Runny Nose Sore Throat Productive cough | Sore throat Muscle aches | N/A | N/A | N/A |
| **Duration** | 3 | 3 | 3 | N/A | N/A | N/A |
| **Progression** | Worse | No change | Improving | N/A | N/A | N/A |

*Voice EHR: Background Health Information*

The background health information provided by both patients and controls (Table 3) exemplified valuable data which was not captured in the initial demographic data (Table 2). For example, Patient A discussed acid reflux and recent use of histamines, both of which may be connected to voice changes or other respiratory biomarkers.[54-55] Control A described multiple potential sources of chronic voice and/or speech changes, which may confuse an AI model attempting to diagnose an acute condition. These included asthma, anxiety, fatigue, brain fog, and dysautonomia. Control B described various thyroid conditions which have been associated with changes in the voice, and Control C explained a history of multiple sclerosis (also known to impact the voice/speech).[56-58]

**Table 3: Background Health Information:** Transcribed voice EHR from patients and controls.

| | |
|---|---|
| **Prompt:** Please tell us background information about your health before your illness, including past health problems and medications. | |
| Patient A | "Overall, I am very healthy. I have seasonal allergies and occasional acid reflux. I do not take any regular medications other than an occasional medicine for seasonal allergies like an antihistamine or an occasional medication for acid reflux." |
| Patient B | "I have good overall health, no chronic conditions. I do have seasonal allergies for which I take Allegra 60 milligrams twice a day." |
| Patient C | Once in a while I will get some back pains, but I've had history of back surgery. And nerve blocks. I really don't have any other pains. Once I did have a little bit of chest pain, but the doctor had me on telemetry and nothing serious was found. And I haven't been that sick. I've been feeling well. I've gotten better. I got better from everything. I get better. Not only my health, my mental health, but my physical health. I am growing rapidly. I'm making more progress as I believe in my own condition." |
| Control A | "So, when I was a teenager, I started passing out after track meets and always had low blood pressure. And they said that I was just hypotensive, even though I wasn't on blood pressure medications, and that I was hyperventilating, and then said that I had athletically induced asthma. I continued on currently having pain, then I was finally diagnosed with endometriosis and had pain for that, which caused the anxiety disorder, because being in a lot of pain all the time is horrible. When I got into my 30s and symptoms started becoming worse, fatigue, lack of concentration, just chronic pain all over my body, like nerve sensations, passing out, not being able to do exercise, total chronic fatigue, and I would stand up from a chair at work and I would just instantly blackout. So, that took me to 2010 to finally be diagnosed via a sweat test and a tilt table test, but I had POTS syndrome and dysautonomia. But they never did anything about it other than put me on meds. They never tried to get to the base of it and said that I was just fine. I wasn't that sick, even though I was in a recliner up to 70% of the day some days as it progressed. Well, then I believe it was in 2014 that I finally got hooked up with Anschutz Center in Colorado, in Aurora, with their neuromuscular clinic, and they actually did complete tests and found out that I have the autoimmune disorder, dysautonomia, POTS, as I had low IVIG levels and issues with my muscle and nerve fibers. And then they found I had a weird antibody or some like blood work that was just odd." |
| Control B | "My health history is I have had atrial fibrillation, which is now cured. I am actively sleeping well, being well, reading well about health. I'm doing everything I can to be a long life for my family, lives to be in their 90s, and I want to have a quality of life at that time also, or perhaps better than they have done. And, let's see, I'm wanting to expand my walking abilities to be able to walk more than I have been after the pandemic. I didn't, haven't walked as much as I would have liked to have done. And, I do have lymphedema in one of my legs, and I work with that, you know, making sure that that continues to stay healthy. The thyroid, I've had that since I was about 18. I was hypothyroid, and then I became hyperthyroid, then I became hypothyroid, and now I'm back to hyperthyroid again, but we've just changed it. So, it's an ongoing, we can never quite get it to be perfect for too very long. I've been looked at for, you know, ultrasounds once a year, and my doctor is a specialist in thyroid disease, and he continues to regulate for me. And, when it's regulated, I feel really good. And, when it's not regulated, I don't feel so great, you know, and I'm quite as sharp or as active, or digestion, you know, changes. So, but, so, and the cancer, I had uterine cancer, but we caught it, and it was grade one, stage one, it was 14 years ago." |
| Control C | "I have breast cancer, stage 1, I've had for 5 years, it's been remission. I also have multiple sclerosis; it's been remission for about 20 years. Both is under control; I have minor symptoms from both. And I was not on any drugs for the cancer or didn't have to get chemo or radiation. It was at the beginning stages of the cancer. And the MS, I have managed to keep it under control by good diet, exercise regularly, and trying to be as stress free as possible." |

*Voice EHR: Longitudinal Illness Descriptions*

Verbal illness descriptions provided not only longitudinal symptom progression but also extensive use of qualifiers ("moderate", "very") that quantify severity or other relationships between signs/symptoms. Additionally, the data contained several instances of patient-initiated interventions within the illness window that could potentially account for fluctuations in audio data. Examples included gargling with saline, vitamin C supplements, and over-the-counter (OTC) medication.

**Table 4:** Example of Current Illness Information from 3 patients.

| | |
|---|---|
| **Prompt:** In as much detail as possible, please tell us how your illness has developed from the time when you first noticed the first symptoms until now. Include any medications you took (like Tylenol) or steps you use to reduce your symptoms. Please use words/phrases like "on the first day", "in the morning", "then", "after that" and use descriptive words like "mild", "severe". No detail is too small. | |
| Patient A | "My symptoms started about three to four days ago. I started to have a slight sore throat and a mild dry cough. I also had a slight headache at that time, but it has since resolved, period. Over the next few days, I have had worsening dry cough and a mild to moderate sore throat, period. My sore throat has remained about the same, but my cough has worsened. I have not felt the need to take medications for my symptoms up to this point, other than I've tried to increase my hydration and increase my sleep." |
| Patient B | "On day one, symptoms started in the afternoon with voice hoarseness, sinus and nasal congestion. By day two, the throat was still hoarse but also sore at this time with increased congestion and a headache. I used ibuprofen and Tylenol for the sore throat pain and the headache. On day three, I had increased congestion, both sinus and nasal, and my lymph nodes were swollen. The sore throat was worse and my voice was only at a whisper and still had a headache. I continued to use Tylenol and ibuprofen. Ibuprofen assisted with the throat pain but did not completely eliminate it. I did do a COVID test. On day three, that came back negative for COVID. Day four, pretty much the same as day three. Throat still sore, no real improvement. Headache and lots of sinus and nasal congestion." |
| Patient C | Let's start talking. Okay. My health is, I guess it's okay. I've been under the weather this week a little bit with a sore throat and with a little bit of coughing and bringing up some sputum, but it's getting much better [extracted from first prompt]. And this past week, I started with having a stuffy nose and a sore throat. And so I started taking, I thought maybe it could be related to allergies. I started taking some over-the-counter medication for day and for night for cold and flu-type symptoms. And that seemed to help. And that's about all. And I drank vitamin C. I did a little gargling with saline. And that's it. |

*Voice EHR: Voice Changes*

Initial viability of voice EHR is further supported by audio data from individuals reporting vocal changes (Table 5). Patients A and B described voice changes due to illness, which can be linked to conventional sound data, thereby ensuring that these changes are considered separately from irrelevant voice/speech anomalies due to lifestyle, recording quality, or other factors. Control A reported voice and speech changes due to dysautonomia, including voice cracks and difficulty speaking coherently. Control B mentioned two separate voice changes due to Atrial fibrillation and hyperthyroidism. Finally, Control C described voice changes due to multiple sclerosis (MS). Each of these voice/speech irregularities could be falsely predicted as an infection or other new, emerging condition. This type of information is not captured in existing datasets.

**Table 5:** Examples of Self-Identified Changes in Voice from three patients.

| | |
|---|---|
| **Prompt:** Please tell us if you or anyone else has noticed any recent changes in your voice or speech. These should be changes that have started around the same time as your illness, not any chronic long-term changes. | |
| Patient A | "My voice has become more raspy and deeper." |
| Patient B | "I did notice a big voice change. In fact, that was the first real symptom on day one, was having a hoarse voice. By day two, it was even more hoarse. And as the day went on, that's when my throat began to get more painful. And by day 3, my voice was at a complete whisper. Today is day 4." |
| Patient C | "I have not noticed any changes in my speech pattern. I'm bilingual. Sometimes I speak in Spanish to my Spanish family and sometimes I speak in English, so I haven't had any problems." |
| Control A | "So a lot of people say that my brain fog is worse due to dysautonomia, my voice gets cracky at times and I search for words and have a hard time pronounciating words that I used to pronounce fine before this." |
| Control B | "Yeah, I think I noticed, I don't have AFib anymore because I had the surgery, but I think I noticed a change in my voice when the AFib started. And I also noticed changes in my voice when my thyroid is active. You can hear it in my voice today, actually. And it affected my singing voice, too, you know, whenever it was going on. I used to have a beautiful singing voice. And with the development of that AFib and this thyroid disease, I think I noticed a big change in my voice, kind of, you know, so. But it sounds more raspy and more irritated, you know, instead of clear and strong." |
| Control C | "With the MS, sometimes the voice is not as strong, it gets a little low on occasions when you're tired or fatigued, sometimes the voice gets a little low because the air can't push up to your diaphragm properly to make the voice sound strong or as clear as usual. So that's the only change with the voice is due to the MS, the cancer, that has no change in the voice at all." |

*Voice EHR: Other Information (Free Response)*

The final Voice EHR prompt was used to capture information regarding other aspects of the patient's life or health which they felt were important and may have impacted their current illness. For example, Patient B mentioned a time just before the illness when cleaning products evoked similar symptoms. Control C talked about residual post-operative pain.

**Table 6:** Example Additional Health Information from three patients.

| | |
|---|---|
| **Prompt:** Is there anything else you think may be affecting your health that you would like us to know? For example, you can tell us about your employment or your lifestyle habits. | |
| Patient A | Checked box indicating there is nothing else they would like to share. |
| Patient B | "On day one I was in a home where there was a cleaning lady and when I first walked in the smell of the cleaning product was so strong that I instantly started to cough and felt some issues." |
| Patient C | Checked box indicating there is nothing else they would like to share. |
| Control A | Checked box indicating there is nothing else they would like to share. |
| Control B | Checked box indicating there is nothing else they would like to share. |
| Control C | "I do have minor, minor effects from both the MS as well as the breast cancer. The breast cancer, I just had pain from the site of the surgery because the tumors were taken out of the right breast and the lymph nodes, a couple of tumors in the lymph nodes. So they managed to get all of the tumors out of the breast and the couple that was in the lymph nodes. And so the only effects I have from that is the pain from the surgery, occasionally I'll get a sharp pain where the surgery was, but that is to be expected, especially when I do a lot." |

## 5. Discussion

Results of this study show that semi-structured Voice EHR data may have equal or additional clinical value compared to manually input data (e.g., multiple choice, short answer). This was true in over 80% of cases, even without considering features such as the correlated conventional acoustic data (i.e., vowel phonation, rainbow passage) or the patient-reported data on other circumstances which may have impacted overall health. The creation of a "voice EHR" system introduces numerous potential benefits to the clinical AI space, particularly in settings (i.e., low- and middle- income countries) without a developed health records system to consistently provide detailed longitudinal data for digital health technologies.

### 5.1 Training Data for Clinical AI

The use of voice EHR as training data for AI models may overcome multiple barriers to the safe deployment of such tools for low-resource settings. EHR-driven AI technologies developed in high-income settings may not provide optimal support to medical decision making in resource-constrained settings where the data may be incomplete, incorrect, or "low tech." While gold-standard annotations like lab results are not collected, prompts which were co-designed by healthcare workers and data collection partnerships with clinics will help ensure the viability of voice EHR.

The HEAR application facilitates the rapid collection of "voice EHR" data in a user-friendly way, without 1) requiring time-consuming and error-prone text data entry on the part of the individual, and 2) enforcing a rigid, pre-defined data schema found in traditional EHR, which may limit the incorporation of information which the patient considers to be important. Furthermore, the process of creating a "voice EHR" may be useful to healthcare workers. In the future, transcribed audio may serve as an accompaniment to clinical notes, reducing the redundancy often associated with data collection and potentially enhancing clinical workflows.

With the introduction of text-sound correlates, Voice EHR may additionally compensate for sources of confusion that are often found in clinical data through "biomarker reinforcement." Even if participants provide incoherent/incomplete data in terms of semantic meaning, the HEAR application still captures voice and breathing data which may independently contribute to the robustness of the data. For example, lapses in patient memory, incomplete notes from healthcare workers, or information reported in colloquial terminology may compromise the value of language data, but acoustic features from the voice may be unaffected in these scenarios. The converse may also be true, in which transcripts of patient-reported health information still provide usable data despite background noise or recording errors (e.g., the device was held too far from the mouth).

For cases in which both modalities are viable, the use of voice/sound data in combination with transcribed health information may capture a more comprehensive composite of diseases with diverse phenotypes, particularly at the time of presentation. Sound data may contribute biomarkers for certain diseases which would not currently be captured in clinician notes. Ultimately, self-reported multimodal audio data expands upon basic health information traditionally used for digital health systems and may allow AI models to better consider chronic conditions, voice changes, speech patterns, word choices indicating mood/sentiment, potential exposures, behavioral influences, and specific disease progression. Compared to similar methods like ambient listening, semi-structured Voice EHR may also reduce the variability of multimodal audio data, potentially enabling machine learning modelling from a smaller sample size. This methodology may also reduce AI biases against clinics/healthcare environments which do not engage in conventional workflows or styles of patient interaction (which could reduce the value of ambient listening in these settings).

## 5.2 Limitations

Implementation of the voice EHR data collection process has presented multiple challenges which must be overcome for adaptation at scale. Prompts for semi-structured data collection, particularly in uncontrolled settings, must be optimized to ensure that patients are easily able to complete the tasks correctly. In the initial voice EHR dataset, there were numerous incomplete samples containing only the initial text survey (no recorded audio). There were also cases in which participants miscategorized themselves as controls - potentially due to unclear criteria - resulting in missing data. Clearer instructions with example videos will be included in future versions of the application. The dataset must also be expanded to ensure access to (1) diverse participants from different demographic subpopulations and (2) data from a broader range of illnesses. The current dataset was mainly collected at a hospital or in the home. However, the highest volume of data for some types of disease (e.g., respiratory infections) might be found in primary/urgent care settings, which may explain the imbalance between chronic and acute conditions. Moreover, this data contains only English speakers, and further study is needed to understand how different languages, levels of literacy, accents, or other linguistic nuances may affect the data transcription process. Finally, in low-bandwidth areas, the simultaneous capture of voice and other modalities like vital signs was time-consuming, posing questions about the scalability of data collection.

## 5.3 Future Work

Future work will mainly involve dataset expansion to additional sites/settings, including tropical disease hospitals in Vietnam and primary/urgent care centers in the United States, enhancing the overall diversity of the data. Moreover, a privacy-aware, patient-controlled option to create a time-series voice EHR may be introduced to collect personalized control data from participants and run longitudinal studies on how changes in voice/speech/language may prognose future health challenges. Future work will also involve the development of AI models which use Voice EHR data to perform specific clinical tasks, such as diagnosis of respiratory conditions or the prediction of hospital admission based on health status in the emergency room.

## 6. Conclusion

This report demonstrates that multimodal audio data can serve as a safe, private, and equitable foundation for new AI models in healthcare. Voice EHR may offer a proxy for detailed time-series data only found in high-resource areas, while simultaneously providing voice, speech, and respiratory data to compliment patient-reported information. Ultimately, AI models trained on voice EHR may be used in the clinic and home, supporting patients in hospital "deserts" where healthcare is not readily accessible. While challenges remain, this work highlights the rich information potentially contained in voice EHR.

**Disclosures/Conflicts of Interest:**
The authors declare no competing non-financial interests but the following competing financial interests. NIH may own intellectual property in the field. NIH and BJW receive royalties for licensed patents from Philips, unrelated to this work. BW is Principal Investigator on the following CRADA's = Cooperative Research & Development Agreements, between NIH and industry: Philips, Philips Research, Celsion Corp, BTG Biocompatibles / Boston Scientific, Siemens, NVIDIA, XACT Robotics. Promaxo (in progress). The following industry partners also support research in CIO lab via equipment, personnel, devices and/ or drugs: 3T Technologies (devices), Exact Imaging (data), AngioDynamics (equipment), AstraZeneca (pharmaceuticals, NCI CRADA), ArciTrax (devices and equipment), Imactis (Equipment), Johnson & Johnson (equipment), Medtronic (equipment), Theromics (Supplies), Profound Medical (equipment and supplies), QT Imaging (equipment and supplies). DAC was supported by the Pandemic Sciences Institute at the University of Oxford; the National Institute for Health Research (NIHR) Oxford Biomedical Research Centre (BRC); an NIHR Research Professorship; a Royal Academy of Engineering Research Chair; the Wellcome Trust funded VITAL project (grant 204904/Z/16/Z); the EPSRC (grant EP/W031744/1); and the InnoHK Hong Kong Centre for Cerebro-cardiovascular Engineering (COCHE).

# References

1. Smyrnakis, Emmanouil, et al. "Primary care professionals' experiences during the first wave of the COVID-19 pandemic in Greece: a qualitative study." *BMC Family Practice* 22.1 (2021): 1-10.
2. https://www.goodrx.com/healthcare-access/research/healthcare-deserts-80-percent-of-country-lacks-adequate-healthcare-access
3. Zhang, Xiaoming, et al. "Physician workforce in the United States of America: forecasting nationwide shortages." *Human resources for health* 18.1 (2020): 1-9.
4. Hoyler, Marguerite, et al. "Shortage of doctors, shortage of data: a review of the global surgery, obstetrics, and anesthesia workforce literature." *World journal of surgery* 38 (2014): 269-280.
5. Shin, Philip, et al. "Time out: the impact of physician burnout on patient care quality and safety in perioperative medicine." *The Permanente Journal* 27.2 (2023): 160.
6. Ortega, Marcus V., et al. "Patterns in physician burnout in a stable-linked cohort." *JAMA Network Open* 6.10 (2023): e2336745-e2336745.
7. Pasquini, Mirko. "Mistrustful Dependency: Mistrust as Risk Management in an Italian Emergency Department." *Medical Anthropology* 42.6 (2023): 579-592.
8. OpenAI. "Gpt-4 technical report. arxiv 2303.08774." *View in Article* 2 (2023): 13.
9. Touvron, Hugo, et al. "Llama 2: Open foundation and fine-tuned chat models." arXiv preprint arXiv:2307.09288 (2023).
10. Li, Chunyuan, et al. "Llava-med: Training a large language-and-vision assistant for biomedicine in one day." *Advances in Neural Information Processing Systems* 36 (2024).
11. https://sites.research.google/med-palm/
12. Celi, Leo Anthony, et al. "Sources of bias in artificial intelligence that perpetuate healthcare disparities—A global review." *PLOS Digital Health* 1.3 (2022): e0000022.
13. Yang, Rui, et al. "Disparities in clinical studies of AI enabled applications from a global perspective." *NPJ Digital Medicine* 7.1 (2024): 209.
14. Jayatilleke, Kushlani. "Challenges in implementing surveillance tools of high-income countries (HICs) in low middle income countries (LMICs)." *Current treatment options in infectious diseases* 12 (2020): 191-201.16.
15. Le, Austin, Benjamin H. Han, and Joseph J. Palamar. "When national drug surveys "take too long": An examination of who is at risk for survey fatigue." *Drug and alcohol dependence* 225 (2021): 108769.
16. Jeong, Dahyeon, et al. "Exhaustive or exhausting? Evidence on respondent fatigue in long surveys." *Journal of Development Economics* 161 (2023): 102992.
17. Tracy, John M., et al. "Investigating voice as a biomarker: deep phenotyping methods for early detection of Parkinson's disease." *Journal of biomedical informatics* 104 (2020): 103362.
18. Suppa, Antonio, et al. "Voice in Parkinson's disease: a machine learning study." *Frontiers in Neurology* 13 (2022): 831428.
19. Tougui, Ilias, Abdelilah Jilbab, and Jamal El Mhamdi. "Machine learning smart system for Parkinson disease classification using the voice as a biomarker." *Healthcare Informatics Research* 28.3 (2022): 210-221.
20. Fagherazzi, Guy, et al. "Voice for health: the use of vocal biomarkers from research to clinical practice." *Digital biomarkers* 5.1 (2021): 78-88.
21. Chintalapudi, Nalini, et al. "Voice Biomarkers for Parkinson's Disease Prediction Using Machine Learning Models with Improved Feature Reduction Techniques." *Journal of Data Science and Intelligent Systems* (2023).
22. Asim Iqbal, Mohammed, Krishnamoorthy Devarajan, and Syed Musthak Ahmed. "An optimal asthma disease detection technique for voice signal using hybrid machine learning technique." *Concurrency and Computation: Practice and Experience* 34.11 (2022): e6856.
23. Idrisoglu, Alper, et al. "COPDVD: Automated Classification of Chronic Obstructive Pulmonary Disease on a New Developed and Evaluated Voice Dataset." *Available at SSRN 4713043* (2024).
24. Raju, Nidhin, D. Peter Augustine, and J. Chandra. "A Novel Artificial Intelligence System for the Prediction of Interstitial Lung Diseases." *SN Computer Science* 5.1 (2024): 143.

25. Borna, Sahar, et al. "A Review of Voice-Based Pain Detection in Adults Using Artificial Intelligence." *Bioengineering* 10.4 (2023): 500.
26. Saghiri, Mohammad Ali, Anna Vakhnovetsky, and Julia Vakhnovetsky. "Scoping review of the relationship between diabetes and voice quality." Diabetes Research and Clinical Practice 185 (2022): 109782.
27. Bensoussan, Yael, et al. "Artificial intelligence and laryngeal cancer: from screening to prognosis: a state of the art review." *Otolaryngology–Head and Neck Surgery* 168.3 (2023): 319-329.
28. Bensoussan, Yaël, Olivier Elemento, and Anaïs Rameau. "Voice as an AI Biomarker of Health—Introducing Audiomics." *JAMA Otolaryngology–Head & Neck Surgery* (2024).
29. Ritwik, K. V. S., Kalluri, S. B., & Vijayasenan, D. COVID-19 Patient Detection from Telephone Quality Speech Data. Preprint at *arXiv* https://doi.org/10.48550/ARXIV.2011.04299 (2020).
30. Usman, Mohammed, et al. "Speech as a Biomarker for COVID-19 Detection Using Machine Learning." *Computational Intelligence and Neuroscience* 2022 (2022).
31. Verde, L. et al. Exploring the Use of Artificial Intelligence Techniques to Detect the Presence of Coronavirus Covid-19 Through Speech and Voice Analysis. *IEEE Access* **9**, 65750–65757 (2021).
32. Verde, L., de Pietro, G., & Sannino, G. Artificial Intelligence Techniques for the Non-invasive Detection of COVID-19 Through the Analysis of Voice Signals. *Arabian Journal for Science and Engineering* https://doi.org/10.1007/s13369-021-06041-4 (2021).
33. Bhattacharya, Debarpan, et al. "Analyzing the impact of SARS-CoV-2 variants on respiratory sound signals." *arXiv preprint arXiv:2206.12309* (2022).
34. Alkhodari, M., & Khandoker, A. H. Detection of COVID-19 in smartphone-based breathing recordings: A pre-screening deep learning tool. *PLOS ONE* **17**(1), 1–25 (2022).
35. Han, J. et al. (2022). Sounds of COVID-19: exploring realistic performance of audio-based digital testing. *Npj Digital Medicine* **5**(1), 16 (2022).
36. Suppakitjanusant, Pichatorn, et al. "Identifying individuals with recent COVID-19 through voice classification using deep learning." *Scientific Reports* 11.1 (2021): 19149.
37. Anibal, James, et al. "Omicron detection with large language models and YouTube audio data". Medrxiv preprint medrxiv: 2022.09.13.22279673 (2024).
38. Deshpande, Gauri, Anton Batliner, and Björn W. Schuller. "AI-Based human audio processing for COVID-19: A comprehensive overview." *Pattern recognition* 122 (2022): 108289.
39. Subirana, Brian, et al. "Hi sigma, do i have the coronavirus?: Call for a new artificial intelligence approach to support health care professionals dealing with the covid-19 pandemic." arXiv preprint arXiv:2004.06510 (2020).
40. Brown, Chloë, et al. "Exploring automatic diagnosis of COVID-19 from crowdsourced respiratory sound data." *Proceedings of the 26th ACM SIGKDD international conference on knowledge discovery & data mining*. 2020.
41. Despotovic, Vladimir, et al. "Detection of COVID-19 from voice, cough and breathing patterns: Dataset and preliminary results." *Computers in Biology and Medicine* 138 (2021): 104944.
42. Coppock, Harry, et al. "End-to-end convolutional neural network enables COVID-19 detection from breath and cough audio: a pilot study." *BMJ innovations* 7.2 (2021).
43. Coppock, Harry, et al. "Audio-based AI classifiers show no evidence of improved COVID-19 screening over simple symptoms checkers." *Nature Machine Intelligence* 6.2 (2024): 229-242.
44. Heaven, Will Douglas. "Hundreds of AI tools have been built to catch covid. None of them helped." *MIT Technology Review. Retrieved October* 6 (2021): 2021.
45. Bhattacharya, Debarpan, et al. "Coswara: A respiratory sounds and symptoms dataset for remote screening of SARS-CoV-2 infection." *Scientific Data* 10.1 (2023): 397.
46. Triantafyllopoulos, Andreas, et al. "COVYT: Introducing the Coronavirus YouTube and TikTok speech dataset featuring the same speakers with and without infection." *arXiv preprint arXiv:2206.11045* (2022).
47. Awan, Shaheen N. "The effect of smoking on the dysphonia severity index in females." *Folia phoniatrica et Logopaedica* 63.2 (2011): 65-71.
48. https://datascience.nih.gov/strides

49. Radford, Alec, et al. "Robust speech recognition via large-scale weak supervision." International Conference on Machine Learning. PMLR, 2023.
50. Fairbanks, G. (1960). Voice and articulation drillbook, 2nd edn.
51. Nam, Yunyoung, Bersain A. Reyes, and Ki H. Chon. "Estimation of respiratory rates using the built-in microphone of a smartphone or headset." *IEEE journal of biomedical and health informatics* 20.6 (2015): 1493-1501.
52. Anibal, James, et al. "Transformers for rapid detection of airway stenosis and stridor." *medRxiv* (2024): 2024-10.
53. https://platform.openai.com/docs/models
54. Abaza, Mona M., et al. "Effects of medications on the voice." Otolaryngologic Clinics of North America 40.5 (2007): 1081-1090.
55. Vashani, K., et al. "Effectiveness of voice therapy in reflux-related voice disorders." Diseases of the Esophagus 23.1 (2010): 27-32.
56. Junuzović-Žunić, Lejla, Amela Ibrahimagić, and Selma Altumbabić. "Voice characteristics in patients with thyroid disorders." *The Eurasian journal of medicine* 51.2 (2019): 101.
57. Stogowska, Ewa, et al. "Voice changes in reproductive disorders, thyroid disorders and diabetes: a review." *Endocrine Connections* 11.3 (2022).
58. Feijó, Adriana Vélez, et al. "Acoustic analysis of voice in multiple sclerosis patients." *Journal of Voice* 18.3 (2004): 341-347.